\documentclass[showpacs,preprintnumbers,amsmath,amssymb]{revtex4}
\usepackage{graphicx}
\begin{document}
\preprint{IFF-RCA-09-07}
\title{Life originated during accelerating expansion in the multiverse}

\author{Pedro F. Gonz\'{a}lez-D\'{\i}az}
\affiliation{Colina de los Chopos, Centro de F\'{\i}sica ``Miguel
A.
Catal\'{a}n'', Instituto de F\'{\i}sica Fundamental,\\
Consejo Superior de Investigaciones Cient\'{\i}ficas, Serrano 121,
28006 Madrid (SPAIN)}

\date{\today}
\begin{abstract}
It is argued that all notions associated with the origin of life
should be related with the participatory anthropic principle of
Wheeler and must be extended into the realm of the multiverse.
Also discussed is the notion that life can only be possible in a
given universe during a finite period along which such a universe
expands in an accelerated fashion. We advance finally the idea
that life, cosmic accelerated expansion and quantum theory are
nothing but three distinct faces from a single, unique coin which
describes the physical reality.
\end{abstract}

\pacs{04.20.Gz , 04.62.+v}

\maketitle

{\bf 1.} The idea that life and cosmology are intimately linked to
one another is not new [1]. Neither is at all recent a set of
assumed likely connections between the notion of life and the
basic principles of quantum mechanics [2]. On the other hand, the
current period of accelerating expansion of the universe could be
related to a state where the universe would adopt a quantum
mechanical behavior (see later) and therefore that period might
also be most straightforwardly related with the emergence of life
in the universe [2]. All of such questions are the subject of a
new scientific discipline which could most naturally be dubbed
"astrobiology", an activity which is flourishing in new research
centers spread throughout the entire world. The aim of this paper
is nevertheless more akin to what can be rather denoted as
biocosmology, which would be hypothesized to be a branch of
cosmology making use of the above biological and cosmological
ideas together with the anthropic principles and the dark and
phantom energy help, in the extent that this paper actually aims
at presenting the hypothesis that for all notions related with the
origin of life and the Wheeler's participatory anthropic principle
[3] to become effective, they should be extended into the realm of
the multiverse.

\vspace{1cm}

{\bf 2.} It appears quite a widespread accepted opinion that the
origin of life is a cosmological problem [1]. In order for life to
be an operative concept in this way two conditions are
simultaneously required to hold: the formation of self-replicating
long molecules and aminoacids and the synthesis of conveniently
folded proteins made of out from such aminoacids. However, if we
assume these two conditions to be fulfilled as a consequence from
the evolution of the universe, then we are confronted with two big
problems. On the one hand, such as it was many times stressed by
Hoyle [4], the probability that self replicating molecules able to
support life had been formed at any place of the universe is
similar to that a tornado has of being able to mount a Boeing 747
out from the materials of a junk-yard. On the other hand, the
well-known Levinthal paradox [5] makes it sure that a
supercomputer which was based on plausible physical-chemical and
spectroscopic rules (such as internal hindered rotation, bending
or wagging vibrations, etc) would take $10^{127}$ years in finding
the native (active for life) configuration of a protein made of
some 100 aminoacids, which is properly folded and has a suitable
biological behavior. It follows that during its entire evolution
until now the universe would only allow for an extremely tiny room
for life to be created anywhere on it.

It has been believed during many years that only the smallest
particles or objects show a quantum-mechanical behavior. However,
recent years have witnessed the emergence of the idea that such a
belief is no longer valid in the realm of the accelerating current
cosmology. In fact (i) if the present universe is filled with
phantom energy (such as it appears to be most supported by
astronomical data [6]), then the larger the universe size the
greater its energy density and therefore a sharper
quantum-mechanical behavior should be expected to be manifested
for the current universe as far as it rapidly expands with time,
tending to a true singularity when the size of the universe and
its energy density are both simultaneously infinity, at the big
rip [7]. On the other hand and quite more importantly, (ii) it has
been recently shown [8] that the ultimate cause for the current
speeding-up of the universe is a universal quantum entanglement
and that one should expect that the very existence of the universe
implied the violation of the Bell's inequalities and hence the
collapse of the superposed cosmic quantum state into the universe
we are able to observe, or its associated complementarity between
cosmological and microscopic laws, and any of all other aspects
that characterize a quantum system as well. Actually, the
formation of molecules able to self replicate is by itself a
quantum process. The billions of smaller molecules in the
primordial soup collided and quantum-mechanically formed trillions
of new molecules throughout quantum processes. In any event, the
probability that these random collisions would produce a molecule
able to self replicate is tiny actually, so tiny that such a
molecule could never have been formed on Earth or on billions of
planets alike [4]. Likewise, protein folding is also a process
governed by quantum rules and describable by a wave equation that
contains a power-law potential which can be expressed in terms of
an order parameter expressible as a scalar field that jointly
represents the set of all internal motions of the molecules along
its normal coordinate modes [9]. A dependence of such a potential
with temperature allows us to express the protein folding process
as described by means of a mechanism of spontaneous symmetry
breaking, the symmetry being the number of contacts among
hydrophobic groups in the protein [9]. In any case the probability
for the above whole process leading to the emergence of life to
occur in a single universe is very small really.

It follows that the generation of life in a single universe like
ours is an extremely unlikely process, in spite of such a single
universe being a quantum-mechanical system without any classical
analog and all physical biological processes leading to life have
a deep quantum character. Even though there were yet unknown
processes that linked protein folding with the creation of
molecules able to self replicating in such a way that once such
molecules are synthesized the required protein folding process
would automatically take place, the creation of life in a single
universe would still be extremely unlikely.

Anthropic principles correspond to a notion that is somehow
against such a conclusion. In particular, the Wheeler notion of
the participatory principle [3], according to which we exist in a
universe which creates itself along a self-reference process. Of
course, this idea has a quantum-mechanical origin as well, and
predicts the existence of observers who are by themselves able to
create all the physical reality that they are able to observe,
even the Big Bang and themselves. Thus, rather than intelligent
observers being created by the universe, what matters here is a
universe which is created and evolved by the observers or at least
an entity where one does not know what was before, the hen or the
egg.

\vspace{1cm}

{\bf 3.} Another crucial notion is that of Boltzmann brains [10].
Perhaps ours is a typical civilization [10,11] that was created by
some random fluctuation from vacuum and its condition of
mediocrity [11] (typicalness) created the universe we are able to
observe or imagine. Such a solution to the problem of unlikeness
is actually not a solution because the Boltzmann spontaneous
fluctuation is a process which is also extremely unlikely to occur
along the life of the universe from the big bang until now. It has
been hitherto said that the probability that molecules able to
self replicate be synthesized and that for even the smallest
proteins to properly fold into their native structure, are
actually tiny. Even so, strictly speaking, such probabilities are
not exactly zero. Therefore, since everything that may happen with
whatever small but still nonzero probability actually happens with
real certainty in the realm of the quantum multiverse [12], our
main hypothesis is that molecules able to self replicate and
properly fold must have been immediately synthesized, and hence
life must necessarily have emerged in the context of the quantum
multiverse. That happened and it did rapidly. The solution of the
problem of the origin of life in the context of the quantum
multiverse must actually be based on an analysis of conditional
probability rather than just probability. In fact, one can always
be sure that however small can be the probability for protein
folding to occur provided the corresponding biological molecules
able to self replicate have already been synthesized, it will not
be strictly zero  and therefore the whole process for originating
life in the quantum multiverse must be true certainty actually.

However, for that to become physically feasible in the cosmic
realm we are dealing with, it ought to conform to both the Wheeler
participatory anthropic principle and a civilization able to be
typical with respect to the universe we know; that is to say, if
we extend the notion of typicalness [10], and hence of mediocrity
[11], to the whole multiverse, a civilization which is typical in
a given universe, would be so also in the whole multiverse, that
meaning that either the typical observers can in someway observe
the universes they are not living in or that such universes do not
actually physically exist, at least from a participatory physical
standpoint principle.

Physical space-time connections between two universes of the
multiverse through which the observers can retrieve some relevant
physical information from these two universes can only be achieved
by means of Lorentzian wormholes with relative speeds between
their mouths at all unspecified. The latter feature expresses the
otherwise mutual independence between the space-times of the
universes that form up the multiverse, and makes it impossible to
establish any kind of simultaneity among distinct civilizations
potentially living in different universes. Therefore, even though
life is originated almost immediately in the whole context of the
multiverse, it can only be realized in just one universe if we
want observers to be typical, no matter whether they are able to
perceive just one universe or many through connections by means of
wormholes, the second possibility being more probable certainly.

\vspace{1cm}

{\bf 4.} The great scientist and scientific divulger Carl Sagan
used to declare [13] that we all were somehow present in the
primeval Big Bang and became later what has been dubbed as powder
of stars. If life actually is an endeavor of the whole multiverse
rather than a matter that concerns particular universes, then the
Sagan's idea had to be actually extended to the context of the
multiverse. Instead of his declared observations, one could well
say that we all were somehow present at the moment in which the
whole multiverse was created, that is to say eternity, quite
likely.

Panspermia is an ever credit-gaining theory which shifts the
origin of life on Earth from Earth itself to the single
cosmological context [14]. Etymologically, it means seeds
everywhere, so expressing the idea that the seeds of life are
spread in a rather homogeneous form throughout the entire
universe, and that such seeds once reached the Earth where they
developed into the known living beings. If life is a matter
concerning the whole extent of the multiverse, then one had to
replace the notion of panspermia for that of what could be dubbed
as holospermia. Etymologically, holospermia would mean {\it seeds
in the wholeness} and would express instead the idea that the
seeds of life were spread throughout the whole multiverse in our
remote past, and that such seeds once reached our own universe,
possibly through a wormhole.

It was Sir Fred Hoyle who coined the term panspermia for his
cosmic theory for the origin of life on Earth [14]. It is
possible- and actually claimed by the notion of holospermia- that,
rather than being present at the Big Bang, we were all originated
in the set of all universes making up the multiverse. Actually, if
as stressed many times by Hoyle himself [4], the probability for
life to have been spontaneously generated in our own universe is
extremely tiny, and that on the contrary, it becomes full
certainty in the set of all universes making up the multiverse,
then the probability for holospermia to be responsible for the
origin of life in our universe is by far much bigger than for
panspermia making that job.

\vspace{1cm}

{\bf 5.} It has been shown [15] that whereas life cannot be
maintained in the future of a de Sitter or decelerating universe,
it can be extended indefinitely in the case that the universe is
filled with dark energy. We can see also in a rather
straightforwardly way that the latter result keeps being valid
also in the case of a universe filled with phantom energy. In
fact, for a constant equation of state with $w= Const. <-1$, the
condition
\begin{equation}
\frac{dH}{dt}=-\frac{3}{2}(1+w)H^2 ,
\end{equation}
amounts to $H\propto -t^{-1}$, so that the Hawking temperature
will turn out to be expressed as $T_H \propto H \propto -t^{-1}$,
which, in the Dyson's notation implies $q=1$ so preserving the
Dyson requirement [15] and hence eternal endurance in the future
of life in a phantom universe. The emergence of a big rip
singularity in a finite time of the future would at first sight
seem to indicate that there will be a doomsday at that singularity
where life, together with all other physical objects and the laws
of science themselves, will inexorably perish. Intervening
wormholes connecting both sides of the singularity might slightly
-in cosmic terms- delay the final destructive destiny of life, but
even though some living patches would bridge the singularity abyss
getting on the other side, the space-time there is contracting
rather than expanding and hence life would again have its hours
counted by application of the Dyson argument. The only way to be
followed by future civilizations and living beings to try to get
in a future trail getting into eternity would be by using the big
trip connections among an infinite number of universes [16]. Thus,
even in the case that the universes are filled with phantom
energy, life could endure eternally in the realm of an infinite
number of universes.

In what follows we will argue that whereas life will in this way
eternally persist in the future of an ever accelerating universe,
it is bound to be confined at times longer than the coincidence
time in the past of that universe. The reason for that confinement
is simple: since any decelerating equation of state does not allow
life to persist long enough in the future of any evolutive
hypersurface when we trace evolution back to a sufficiently early
time, one would always have a situation which is lifeless before
the coincidence time.

This result would confirm the intuition that ultimately life is
nothing but a property of the accelerated period of the universe,
a period which, on the other hand, is closely related to the deep
quantum-mechanical character of the universe in such a way that it
should somehow be connected with sharp quantum properties such as
entanglement, wave packet reduction and non-locality [8]. In this
way, life, cosmic accelerated expansion and quantum theory are
nothing but three distinct faces from a single, unique coin.

Let us finally briefly consider the issue of life survival in
relation with the second law of thermodynamics in the contexts of
our single universe and the multiverse. We notice that such an
issue can be dealt with by using the following two analogies. On
the one hand, one has the well known Schr\"{o}dinger idea, which
was advanced in his famous book "What is life?", that life is
noting but information (in the Shannon sense) or, in
Schr\"{o}dinger terminology, negative entropy or negentropy; on
the other hand, it has been many times stressed that the
biological process of self-replication is equivalent to
computation, that is to say, it is like a computer. By adopting
these standpoints, one can deduce that in an accelerating
universe, one can see less infinite space rather than more of it.
The bounds of  the observable universe shrink as the space between
objects accelerate and expand as the spaces close because no light
from objects outside a range of 13.7 billion light years - the
time of the birth of the universe - has enough time to reach the
Earth. It follows that entropy in our universe should increase
very quickly in the presence of dark or phantom energy. Moreover,
since entropy increases rapidly in an ever accelerating universe,
a computer could not run forever in an ever accelerating universe
like ours. Therefore, life cannot last forever in the presence of
dark and phantom energy in a single universe.  Clearly, it can
only be in the context of the multiverse that this entropic effect
can be compensated by the opening of an infinite number of
classical (and possibly quantum) information channels which can
ultimately render life itself to last forever. Such classical
information channels would be made of the above alluded
inter-universal wormhole connections. Moreover, a computer
(self-replicating biological system) which can run forever this
way has a potentially infinite amount of memory available. It then
follows that, in the context of the multiverse, every thought will
be destined not to be forgotten and then re-discovered, but rather
to be preserved forever.

In this Letter we have discussed some relations between the
current evolution of our accelerating universe and the origin of
life and intelligent civilizations in the full context of a
multiverse model where the distinct universes are linked to each
other by means of traversable Lorentzian wormholes. Adhering to
the recent view that the emergence of life is a business of the
whole multiverse rather than individual universes, we argue in
favor of the ideas that once life appears in the multiverse it
lasts forever, and that life, the accelerating universe and the
deepest aspects of the quantum theory are nothing but three
distinct faces from a single coin describing the physical reality.
Also favored is the idea that the knowledge which is being
achieved by the civilizations will be accumulated and preserved
forever.

\acknowledgements

\noindent The author thanks Carmen L. Sig\"{u}enza for useful
discussions and the members of the theater group of Medell\'{\i}n,
Spain, who allowed me to do this work in a nice
scientific-artistic working atmosphere. This work was supported by
MEC under Research Project No. FIS2008-06332.

\end{document}